\documentclass[runningheads]{llncs}
\usepackage[T1]{fontenc}
\usepackage[utf8]{inputenc}
\usepackage{amsmath,amssymb,graphicx,tikz,xcolor,hyperref,booktabs,upquote,csquotes,textcomp,multirow}
\usepackage{pifont}
\usepackage{xcolor}
\usepackage{xspace}
\usepackage{enumerate}
\usepackage{caption}
\usepackage{subcaption}

\usetikzlibrary{arrows,decorations.pathmorphing,backgrounds,positioning,fadings,shadings,through,fit,petri,calc}

\newcommand{\badbill}{\tilde{B}}

\newcommand{\node}{v}
\newcommand{\nodea}{v}

\newcommand{\ignore}[1]{}

\newcommand{\true}{\textsc{True}\xspace}
\newcommand{\false}{\textsc{False}\xspace}

\usepackage{comment}

\newcommand{\comb}[2]{\left( \begin{array}{c} #1 \\ #2 \end{array} \right)}

\newcommand{\rand}{\stackrel{{\scriptscriptstyle\$}}{\leftarrow}}
\newcommand{\todo}[1]{{\textcolor{red}{TODO}}\textcolor{red}{(todo: #1)}}

\newcommand{\GameP}{\ensuremath{\mathsf{Game}}}
\newcommand{\Adv}{\ensuremath{\mathsf{Adv}}}
\newcommand{\Adversary}{\ensuremath{\mathcal{A}}}
\newcommand{\Comm}{\ensuremath{\mathsf{COM}}}
\newcommand{\CommSetup}{\ensuremath{\mathsf{COM.Setup}}}
\newcommand{\CommCommit}{\ensuremath{\mathsf{COM.Commit}}}

\newcommand{\ParmsC}{\ensuremath{\text{P}_\text{c}}}
\newcommand{\VSpace}{\ensuremath{\mathcal{V}_\text{c}}}
\newcommand{\RSpace}{\ensuremath{\mathcal{R}_\text{c}}}
\newcommand{\CSpace}{\ensuremath{\mathcal{C}_\text{c}}}

\newcommand{\NIZK}{\ensuremath{\mathsf{NIZK}}}
\newcommand{\NIZKSetup}{\ensuremath{\mathsf{NIZK.Setup}}}
\newcommand{\NIZKProve}{\ensuremath{\mathsf{NIZK.Prv}}}
\newcommand{\NIZKVerify}{\ensuremath{\mathsf{NIZK.Vrf}}}

\newcommand{\RelationSpace}{\ensuremath{\mathsf{R_{zk}}}}

 \newcommand{\ParmsZK}{\ensuremath{\text{P}_{\text{zk}}}}

\newcommand{\Qlist}{\ensuremath{\mathsf{QL}}}

\newcommand{\PRF}{\ensuremath{\mathsf{PRF}}}
 
  \newcommand{\PRFKGen}{\ensuremath{\mathsf{PRF.KGen}}}
\newcommand{\PRFEval}{\ensuremath{\mathsf{PRF.Eval}}}

\newcommand{\ParmsTP}{\ensuremath{\text{P}_\text{tp}}}
\newcommand{\TPInitialize}{\ensuremath{\mathsf{Initialize}}}
\newcommand{\TPSlotSecretGen}{\ensuremath{\mathsf{SlotSecretGen}}}

\newcommand{\TPEvidenceGen}{\ensuremath{\mathsf{EvidenceGen}}}
\newcommand{\TPEvidenceVerify}{\ensuremath{\mathsf{EvidenceVrf}}}

\newcommand{\Evidence}{\ensuremath{\text{E}}}
\newcommand{\TPEvidenceSpace}{\ensuremath{\mathcal{E}_{\text{tp}}}}
\newcommand{\TPSKSpace}{\ensuremath{\mathcal{SK}_{\text{tp}}}}
\newcommand{\TPSSSpace}{\ensuremath{\mathcal{SS}_{\text{tp}}}}
\newcommand{\TPVSpace}{\ensuremath{\mathcal{V}_{\text{tp}}}}

\newcommand{\Consistency}{\ensuremath{\mathsf{Trans}}}

\newcommand{\Privacy}{\ensuremath{\mathsf{Priv}}}

\newcommand{\PPPT}{\ensuremath{\mathsf{PPTP}}}

\newcommand{\OracleEGen}{\ensuremath{\mathcal{O}_{\mathsf{EG}}}}
\newcommand{\OracleCorrupt}{\ensuremath{\mathcal{O}_{\mathsf{C}}}}

\newcommand{\CorruptList}{\ensuremath{\mathsf{CL}}}

\newcommand{\PPTPRelation}{\ensuremath{\mathcal{RL}}}
\newcommand{\Vmax}{\ensuremath{\mathsf{v}_{\text{max}}}}

\begin{document}
\title{Transparent Electricity Pricing with Privacy}

\author{Dani\"el Reijsbergen$^{*}$ \and Zheng Yang\orcidID{0000-0001-8610-9936}\thanks{Dani\"el Reijsbergen and Zheng Yang are the Corresponding Authors.},
Aung Maw \and \\ Tien Tuan Anh Dinh \and Jianying Zhou}
\authorrunning{Reijsbergen et al.}
\institute{Singapore University of Technology and Design} 

\maketitle

\vspace{-0.2cm}

\begin{abstract}
Smart grids leverage data from smart meters to improve operations management and to achieve cost reductions. The
fine-grained meter data also enable pricing schemes that simultaneously benefit electricity retailers and users. Our
goal is to design a practical dynamic pricing protocol for smart grids in which the rate charged by a retailer depends
on the total demand among its users. Realizing this goal is challenging because neither the retailer nor
the users are trusted. The first challenge is to design a pricing scheme that incentivizes consumption behavior that leads to lower costs for both the users and the retailer.  The second challenge
is to prevent the retailer from tampering with the data, for example, by claiming that the total
consumption is much higher than its real value. The third challenge is data privacy, that is, how to
hide the meter data from adversarial users. To address these challenges, we propose a scheme in which peak
rates are charged if either the total or the individual consumptions exceed some thresholds. We formally
define a privacy-preserving transparent pricing scheme (PPTP) that allows honest users to detect tampering at
the retailer while ensuring data privacy. We present two instantiations of PPTP, and prove their security. Both protocols use secure commitments and zero-knowledge proofs. We implement and
evaluate the protocols on server and edge hardware, demonstrating that PPTP has
practical performance at scale.  

\end{abstract}

\section{Introduction}
\label{sec:introduction}
\textit{Smart meters}, the building blocks of smart grids, allow electricity retailers to measure their
customers' power use at a frequency that is far beyond the capabilities of traditional meters, which rely on
sporadic manual readouts. One important use case for a higher data frequency is that it enables more accurate and
informative bills. The high measurement frequency also enables advanced electricity \textit{pricing}. Since
the retailer's costs are highest during periods of high network demand, it is often profitable to incentivize
users to spread their demand across the day by charging them different rates during \textit{peak} and
\textit{off-peak} periods. This is ultimately also beneficial for the customers themselves because a retailer
with lower costs can offer more competitive prices.  

Pricing schemes that are based on fixed peak and off-peak rates do not make full use of the smart meter's
potential, which includes the ability to receive up-to-date pricing information that reflects network
conditions observed by the retailer.  To improve on this, \textit{dynamic} pricing schemes
\cite{vardakas2014survey} allow the retailer to anticipate periods of high electricity demand, e.g.,
heatwaves or major sporting events, and incentivize users to shift their loads away from these periods. In
this set-up, the retailer shares pricing information with the meters at the start of each operational cycle (e.g.,
at midnight), which can then be forwarded to a smart home hub that coordinates high-demand activities (e.g.,
charging an electric vehicle) across the household. In return, the meter periodically sends measurements back to the retailer,
and after a number of \textit{operational cycles} -- an operational cycle typically lasts a single day -- the user is presented with an electricity bill that reflects the power usage and prices throughout the cycles. However, prior  
research \cite{allcott2009real,vardakas2014survey} has found that users are uncomfortable with pricing schemes
that they perceive to be overly complex or risky.

Our goal is to design a practical dynamic pricing protocol. This is challenging because neither the retailers nor the
users are trusted. In particular, the first challenge is to design a pricing scheme that
can simultaneously reduce the costs of the retailer and the users -- i.e., that the retailer is able to charge higher
prices during periods when the actual demand is high across the network. The second challenge is to prevent
the retailer from giving wrong information about network-level demand -- i.e., that it cannot tamper with the
aggregate measurements from the meters. The third challenge is to share network-level information without
revealing the users' privacy-sensitive measurements to other users -- i.e., that an honest-but-curious user
cannot learn other users' measurements.

The latter two challenges have been studied in the related literature, e.g., by Shi et al.\
\cite{shi2011privacy} and others \cite{acs2011have,erkin2012private}. However, these works use a
different threat model than the one that is most relevant in our setting. For example, in the work by Shi et al.\
\cite{shi2011privacy} the retailer is not trusted with knowledge of individual users' measurements. Since the
retailer in our model needs the measurements anyway to compute the bills, we can afford a more relaxed trust
assumption, namely that the privacy requirement only applies to honest-but-curious users. Another difference is
that Shi et al.~\cite{shi2011privacy} do not consider strict integrity of the aggregate data -- therefore, in their setting
random noise can be added to measurements, but allowing this in our setting would mean that the retailer can increase
its profits by inflating the aggregate value. In summary, our threat model gives us opportunities to design a
more efficient privacy-preserving protocol, while at the same time necessitating the protection of measurement
integrity.

To address the above challenges, we present a privacy-preserving and transparent pricing
(PPTP) scheme. The scheme addresses the first challenge by charging a peak or off-peak rate depending on the
\textit{actual} network demand. This protects the retailer from the risk of financial
losses during a demand surge while being fully transparent to the user. It is predictable for the
user as network demand under normal circumstances can be predicted with reasonable accuracy given historical
data \cite{mohsenian2010optimal}. To address the second challenge, we require the retailer to share verifiable
proofs about users' measurements. To address the third challenge, we use zero-knowledge range proofs that
do not reveal the underlying measurements, e.g., Bulletproofs \cite{bunz2018bulletproofs}. We formalize the
  security definition of PPTP and present two instantiations: a baseline protocol that shares the full set
  of range proofs with each user, and a more efficient one that uses Merkle trees and which uses concepts from related work on Certificate Transparency \cite{CCS:ChaseM16,laurie2014certificate}. We
  prove the security of both instantiations, and evaluate their performance on a server and on edge devices.
  The results show that PPTP achieves practical performance at scale. 

In summary, we make the following contributions.
\begin{enumerate}
\item We present a dynamic pricing protocol for smart grids that reduces the cost for both the retailer and users.  
\item We define a privacy-preserving and transparent pricing (PPTP) scheme that ensures integrity of the
retailer when computing dynamic prices. PPTP protects confidentiality of the meter data against malicious users. 
\item We instantiate two PPTP protocols and prove their security properties. 
\item We implement the two protocols and demonstrate their performance at scale using a server and Raspberry Pi hardware. 
\end{enumerate}
Throughout this paper, we use the notation $[k]$ to denote $\{1,\ldots,k\}$ for $k \in \mathbb{N}$. A summary of the other notation used in the paper can be found in Table~\ref{tab:notation}.
The structure of the paper is as follows.
In Section~\ref{sec:system_model}, we present our model of the system's cost and price structure. The network model, threat model, and requirements are presented in Section~\ref{sec:security_model}. The two PPTP instantiations are presented in Sections~\ref{sec:baseline}~and~\ref{sec:solution}. We present the experimental results in Section~\ref{sec:implementation} and discuss related work in Section~\ref{sec:related}. Section~\ref{sec:conclusions} concludes the paper.

\begin{table}[!ht]
\vspace{-0.4cm}
    \caption{Summary of the notation used in this paper.}
	\label{tab:notation}
    \begin{center}
    \begin{scriptsize}
    \begin{tabular}{ccccl}
    symbol & \;\; & values & \;\; & \multicolumn{1}{c}{meaning} \\ \toprule
    $n$ & & $\mathbb{N}$ & & number of users \\
    $k$ & & $\mathbb{N}$ & &  number of time periods per operational cycle\\
    $p_{i\,t}$ & & $\mathbb{N}$ & & rate charged to user $i \in [n]$ in period $t \in [k]$ \\
    $B_i$ & & $\mathbb{N}$ & & bill of user $i \in [n]$ \\
    $\alpha_t$ & & $\mathbb{N}$ & & peak rate in period $t \in [k]$ \\
    $\beta_t$ & & $\{0,\ldots,\alpha_t\}$ & & off-peak rate in period $t \in [k]$ \\
    $\gamma_t$ & & $\{0,n\delta_t-1\}$ & & system peak rate threshold in period $t \in [k]$ \\
    $\delta_t$ & & $\mathbb{N}$ & & individual peak rate threshold in period $t \in [k]$ \\
    $y_{i\,t}$ & & $\mathbb{N}$ & & raw measurement of user $i \in [n]$ in period $t \in [k]$ \\
    $x_{i\,t}$ & & $\{0,\delta_t\}$ & & truncated measurement of user $i \in [n]$ in period $t \in [k]$ \\
    $r_{i\,t}$ & & $\RSpace$ & & random secret of user $i \in [n]$ in period $t \in [k]$ \\
    $c_{i\,t}$ & & $\CSpace$ & & commitment of $x_{i\,t}$ \\
    $\pi_{i\,t}$ & & & & zero-knowledge proof that $x_{i\,t} \in [0, \delta_t]$ \\
    $x^{*}_t$ & & $\{0,n\delta_t-1\}$ & & sum of $\min(x_{i\,t}, \delta_t)$ for all $i \in [n]$ in period $t \in [k]$ \\
    $c^*_t$ & & $\CSpace$ & & sum of $c_{i\,t}$ for all $i \in [n]$ in period $t \in [k]$ \\
    $\pi^*_t$ & & & & zero-knowledge proof that $x^*_t \in [0, \gamma_t]$ \\
    \end{tabular}
    \end{scriptsize}
    \end{center}
    \vspace{-0.5cm}
\end{table}

\section{Electricity Pricing}
\label{sec:system_model}
\vspace{1mm}{\noindent \bf Power usage and cost model.} For the meter readings, we consider a single operational cycle
(typically a day) that is divided into $k$ time periods -- e.g., if each operational cycle lasts for one day
and \mbox{$k=24$}, then each time period corresponds to an hour. Let \mbox{$y_{i\,t} \in \mathbb{N}$} be the
reading by the smart meter of customer $i\in [n]$ in period $t \in [k]$ -- note that measurements can be represented as integers through rounding, e.g., a power use of $2.72$ kWh in a period can be represented as `$272$' if measurements are rounded to two decimals. Let $\vec{y}_t =
(y_{1\,t}, \ldots,y_{n\,t})$ denote the readings in $t$. The readings reflect the customer's energy consumption, which
for every period equals the sum of all the \emph{loads} run in that period. A load is an
  electricity-consuming job, typically related to a specific appliance, that can be represented by its
  required duration (in terms of periods) per operational cycle, and its energy use per period. We consider
  two types of loads: \emph{controllable} and \emph{must-run} loads
  \cite{samadi2012advanced,pedrasa2010coordinated}. 
  The customer
is free to choose when to run the controllable loads, but the must-run loads must be executed in their assigned time
periods.
If the electricity prices $p_t$ depend on the time period $t \in [k]$, then the goal of each
customer $i \in [n]$ is to divide her controllable runs over the operational cycle such that her
total electricity bill $B_i = \sum_{t=1}^k p_{t} y_{i\,t}$
is minimal. 

The retailer's goal is to maximize its profit, which is the difference between its revenues $U$ and its
costs $C$. The retailer's total revenues in a cycle equal the sum of the bills of its customers, i.e., $
U(\vec{y}_t, \ldots, \vec{y}_k) = \sum_{t=1}^k \sum_{i=1}^n  p_{t} \, y_{i\,t}$. Its costs 
depend on the network demand, because a higher network demand means that more energy needs to be purchased on the electricity market. In this work, we do not consider other fixed costs such as fees paid to the grid
operators. As a result, the total costs in a cycle are
given by $C(\vec{y}_t, \ldots, \vec{y}_k) = \sum_{t=1}^k C_t(\vec{y}_t)$, where $C_{t}(\vec{y}_t)$ denotes
the total costs in period $t$ as a function of the demand in $t$. 

We make the following two assumptions
about the retailer's cost function. First, the costs may depend on the time period, which captures
fluctuations in energy supply throughout the day, e.g., due to solar panels which require daylight to operate.
Second, the marginal costs
are an increasing function of the total demand $s_t = \sum_{i=1}^n y_{i\,t}$ -- i.e., if the amount
of energy purchased during a period goes up, then so does the cost per unit of energy. This captures the fact
that when demand exceeds the maximum capacity of the power generators, they must resort to limited reserves
\cite{wood2013power} or operate beyond capacity which increases the generators' long-term
maintenance costs~\cite{samadi2012advanced}.

\vspace{1mm}{\noindent \bf Pricing model.}
It follows from the two assumptions above that the retailer has minimal costs if demand is high during periods
with low prices, and the demand within a single period does not become too high.  The retailer
cannot directly control the demand of its customers in each period, but it can \emph{incentivize} them
to perform load balancing through a pricing mechanism that rewards customers for keeping the retailer's costs
low. 

The problem of pricing mechanism design has been studied extensively in the literature~\cite{luh1982load,vardakas2014survey}.
The most common design is \emph{peak load pricing}: for each
period $t \in [k]$ there is a different rate that depends on energy market prices and historical demand in
$t$.  Although such a scheme incentivizes customers to shift controllable loads away from the peak periods,
they still have no incentive to spread demand across the different non-peak periods.
A more refined pricing mechanism is \emph{real-time
pricing with inclining block rates}~(RTPIBR)~\cite{mohsenian2010optimal}, in which the price paid by customer $i \in
[n]$ during period $t \in [k]$ is given by \begin{equation}
p_{i\,t}(\vec{y}_t) = \left\{\begin{array}{rl} \alpha_t & \text{if }y_{i} > \gamma_t, \\ \beta_t & \text{otherwise.} \end{array}\right.
\label{eq:rtp-ibr}
\end{equation}
Here, $\alpha_t$ is period $t$'s penalty rate, $\beta_t$ is period $t$'s normal rate, and $\gamma_t$ is the
customer's usage threshold in period $t$. In words, in addition to the different prices per period, the
customer is also charged a penalty if her demand in a single period is too high. Although this is an
improvement over peak load pricing, it does not take into account the total network demand. Although the
customers are motivated to shift loads over several non-peak periods to avoid the penalty rate in a single
period, they could all shift it to the same non-peak periods, which would lead to unnecessarily high
demand in those periods. 

We propose a pricing scheme that takes the total demand into account. Our
scheme allows users to learn the expected demand from previous operational cycles and allocate their
controllable loads accordingly. At the start of each cycle, the cycle's penalty prices $\alpha_t$, normal
prices $\beta_t$, network demand thresholds $\gamma_t$, and individual demand thresholds $\delta_t$ are fixed
for all $t \in [k]$ and made available to all the entities, e.g., by posting them on a  
public bulletin board. $\delta_t$ represents the maximum contribution of a single user to the network demand.
Without this bound, a handful of heavy-use customers could report an energy use that exceeds
$\gamma_t$, and the peak rate would apply regardless of the consumption of the other customers. 
  The price
  $p_{i\,t}$ paid by user $i$ during period $t$ is then given by 
  \begin{equation} 
  p_{i\,t}(\vec{y}_t)
= \left\{\begin{array}{rl} \alpha_t & \text{ if }\sum_{j=1}^n \min(y_{j\,t},\delta_t) > \gamma_t \text{ or }
y_{i\,t} > \delta_t, \\ \beta_t & \text{ otherwise.} \end{array}\right.
\label{eq:price}
\end{equation}
This is similar to \eqref{eq:rtp-ibr}, except that the penalty rate is charged depending on the total demand
among the retailer's users \textit{and} whether the individual demand exceeds a threshold. In the following, we write $x_{i\,t} = \min(y_{i\,t},\delta_t)$ and \mbox{$x^*_t = \sum_{i=1}^n,
x_{i\,t}$}, and we write $p(y_{i\,t}, x^*)$ instead of $p_{i\,t}(\vec{y}_t)$ because each user's rate only depends on her own measurement and the sum.

\begin{figure}
\centering
    \begin{subfigure}[t]{.32\textwidth}
      \centering
      \includegraphics[width=\linewidth]{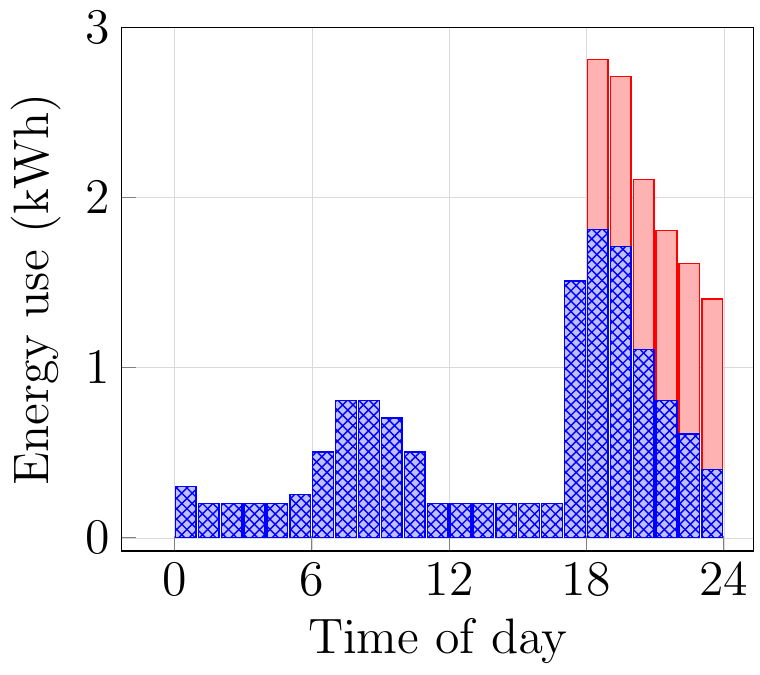}\vspace{-0.15cm}  
      \caption{Na\"ive pricing}
      \label{fig:example-first}
    \end{subfigure}
    \begin{subfigure}[t]{.32\textwidth}
      \centering
      \includegraphics[width=\linewidth]{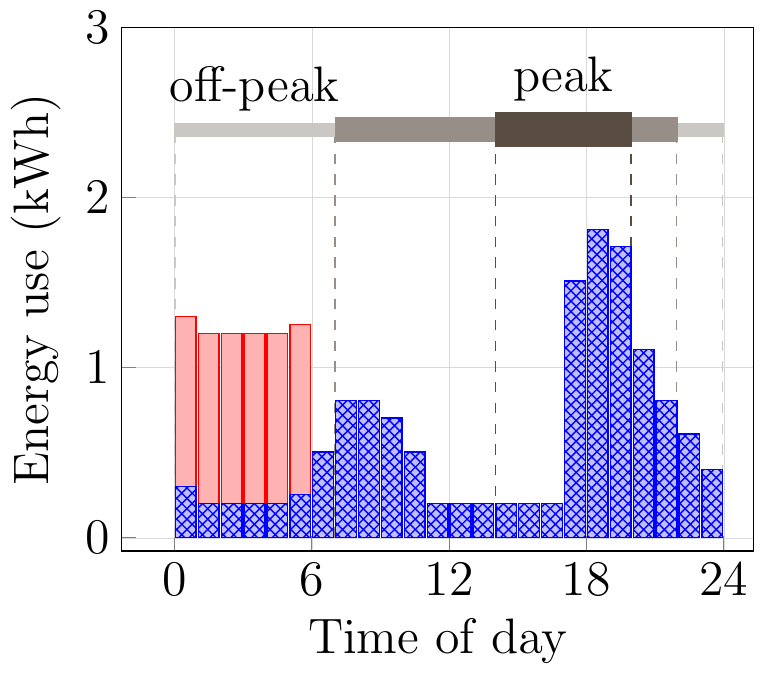}\vspace{-0.15cm}  
      \caption{Peak load pricing}
      \label{fig:example-second}
    \end{subfigure}\vspace{-0.05cm}  
    \begin{subfigure}[t]{.32\textwidth}
      \centering
      \includegraphics[width=\linewidth]{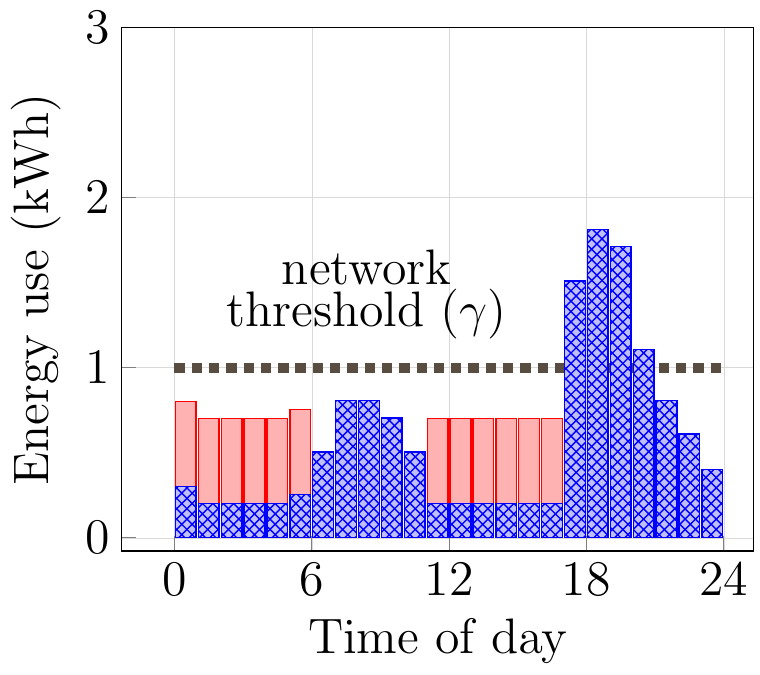}\vspace{-0.15cm}   
      \caption{Our pricing scheme}
      \label{fig:example-third}
    \end{subfigure}
    \caption{Demand over time of an \textit{average} user. Blue and red bars represent demand due to must-run and controllable loads, respectively.}
	\label{fig:example}
	\vspace{-0.5cm}
\end{figure}

\vspace{1mm}{\noindent\bf Example.}
We consider $n$ users who have the same must-run demand
profile every day (taken from \cite{pedrasa2010coordinated}). Each user has a single controllable load ---
namely the charging of an electric vehicle --- which consumes 1 kW for six consecutive hours. We consider
three different usage patterns: (a) all consumers schedule the load at 6PM, (b) all consumers schedule the
load at midnight, and (c) half of the consumers schedule the load at midnight and the other half at noon. These
patterns are shown in Figures~\ref{fig:example-first}, \ref{fig:example-second}, and
\ref{fig:example-third}, respectively.  The retailer's increasing marginal costs imply that its costs are
highest in pattern (a), lower in (b), and lowest in (c).  Using a na\"ive pricing scheme -- i.e., a constant
price -- all three patterns have the same costs for users since they are charged the same rate in all cases. Using peak load pricing~\cite{pedrasa2010coordinated}, in which a peak rate is
charged between 2PM and 8PM and an off-peak rate between 10PM and 7PM, it is no longer optimal for all
consumers to run the load at 6PM. Instead, they may all shift their loads to midnight in order to get the
lowest bills, as shown in
\autoref{fig:example-second}.  
Using RTPIBR, we can impose that consumers also pay the peak rate whenever their demand in a period exceeds
$\gamma = 1 \text{kW}$.  However, although this increases the bill of every user, they are not motivated to  
spread loads over the day as the user still pays a similar rate regardless of whether she
schedules her load at midnight or at noon. Using our pricing scheme in Eq.~\ref{eq:price},  which charges the
peak rate in periods where the network-level demand exceeds $n$kW, the users are motivated to spread their load
to different periods. It can be seen in \autoref{fig:example-third} that both the users and retailer incur
low cost, because the users pay the normal rate, and the retailer has low marginal cost due to a balanced
network demand.

\ignore{
\newcommand{\high}{\text{high}}
\newcommand{\medium}{\text{medium}}
\newcommand{\low}{\text{low}}

\begin{table}[]
    \centering
    \caption{Comparison of retailer and user costs for different load patterns and pricing schemes. }
    \begin{tabular}{c|c|cccc|}
        &  retailer & \multicolumn{4}{c|}{user costs} \\
        timing of controllable load & costs & na\"ive & peak load & RTPIBR & network load \\ \hline
         all users at peak hour (a) & $\high$ & $\medium$ & $\high$ & $\high$ & $\high$\\
         all users at night (b) & $\medium$ & $\medium$ & $\low$ & $\medium$ & $\medium$\\
         half at night, half in afternoon (c) & $\low$ & $\medium$ & $\low$ & $\medium$ & $\low$\\ \hline
    \end{tabular}
    \label{tab:example}
\end{table}
}

\section{System \& Security Model}
\label{sec:security_model}
\begin{figure}[hb!]
\vspace{-0.5cm}
    \centering
	\includegraphics[width=0.85\linewidth]{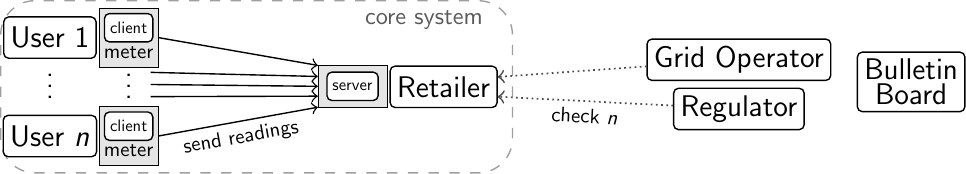}
  \caption{Entities in our system.}
	\label{fig:system_diagram}
	\vspace{-0.4cm}
\end{figure}

\vspace{-0.2cm}{\noindent\bf System model.} 
Figure~\ref{fig:system_diagram} depicts the different entities in our system. There is one power \emph{retailer}
and $n$ \emph{customers} (or users). 
The retailer
purchases electricity from power {generators} on the wholesale electricity market, and pays a fee to one
or more \emph{grid operators} for the use of their transmission and distribution networks. 
We assume that
the regulator and grid operators know $n$, but that
they are unable to observe the individual or total consumption of the retailer's users.\footnote{Although the grid
operators will typically perform their own high-level measurements for system monitoring, e.g., using phasor
measurement units, they will not be able to distinguish between the customers of different retailers in a
small area.} The retailer runs a \textit{server}, and each user owns a smart \textit{meter}, which runs a software
application called a \textit{client} that communicates with the server.

At the end of each period $t \in [k]$, each smart meter $i \in [n]$
sends the raw reading $y_{i\,t}$ to the retailer via a secure and highly available channel. 
At the end of the operational cycle, the retailer uses the same channel to send to each user a
statement containing the total bill $B_i \in \mathbb{N}$, and for each period $t \in [k]$ both $y_{i\,t}$
and the sum $x^*_t$.  
The customer can then verify that $B_i = \sum_{t=1}^k p_{i\,t}(y_{i\,t},x^*_t)
x_{i\,t}$.  However, the customer cannot verify $x^*_t$, which may allow the retailer to charge the peak rate
incorrectly.  

\vspace{1mm}{\noindent\bf Threat model.}
The retailer is adversarial and interested in convincing users of accepting an incorrect bill
$\badbill_i$  such that $\badbill_i > B_i$. 
We assume that the meters are secure, i.e., they produce readings that match the user's true power
consumption -- otherwise, honest users would be able to report this to the regulator. The retailer can  
 collude with a number of \emph{adversarial users} who may generate arbitrary readings. Users can be
adversarial because they have been bribed by the retailer, or they can be Sybil users created by the
retailer. We assume that the regulator would notice if the retailer would create too many Sybils, so this
number is limited in our model.  
Non-adversarial users are honest-but-curious, i.e., they will faithfully follow the
protocol as described, but will attempt to use the information shared with them to learn the readings of other
users.  Finally, there is a public bulletin board which provides immutable and tamper-evident storage, e.g., a
public blockchain. 

Given this threat model, our first goal is allow users to \textit{detect incorrect bills}. In particular, let  $B^*_i$
be the bill for user $i$ if each adversarial user $j$ reports $x_{j\,t} = \delta_t$ in each period $t \in
[k]$. An incorrect bill $\badbill_i$ then satisfies \mbox{$\badbill_i > B^*_i$}. In other words, the adversarial
users should not be able to influence individual bills beyond reporting the maximum. Another goal is to \textit{hide
private readings} $x_{i\,t}$ from other honest-but-curious users. The final goal is \textit{efficiency}, that is, the
protocol can support a large number of users with practical performance.

\subsection{Security Model} 
We let $\kappa$ be the security parameter by $\kappa$, and $\emptyset$ be the empty string. 
When $X$ is a set, $x \rand X$ denotes the action of sampling an element uniformly at random from $X$. %
Let $\|$ denote an operation that concatenates two strings.

We define a privacy-preserving, transparent pricing (PPTP) scheme as one that consists of the four 
algorithms below. 
\vspace{-0.075cm}
\setlength{\leftmargini}{0.1em}
\begin{itemize}%
\setlength\itemsep{0.15em}

\item[]$(\ParmsTP, k_r) \leftarrow \TPInitialize(1^\kappa)$: run by the retailer. It takes as input the
security parameter $\kappa$, and outputs the system parameters $\ParmsTP$ and a secret key $k_r \in
\TPSKSpace$ where $\TPSKSpace$ is the secret key space. 

\item[] $(r_{i\,t})_{i\in[n]} \leftarrow  \TPSlotSecretGen(\ParmsTP,k_r,t)$: run by the retailer. It 
takes as input the system parameters $\ParmsTP$, the secret key $k_r$, and the time slot
$t$, and outputs slot secrets $(r_{i\,t})_{i\in[n]} \in  \TPSSSpace$ for $n$ users, where $\TPSSSpace$ is the
slot secret space. 

\item[] $\Evidence_t \leftarrow \TPEvidenceGen(\ParmsTP,k_r,(x_{i\,t})_{i\in[n]}, t)$: run by the retailer. It
takes as input the system parameters $\ParmsTP$, secret key $k_r$, and the measurements
$(x_{i\,t})_{i\in[n]} \in \TPVSpace$ of each user $i$ for time $t$, and outputs an evidence $\Evidence_t \in
\TPEvidenceSpace$. Here, $\TPEvidenceSpace$ is the evidence space and $\TPVSpace$ is the measurement space. 

\item[] $\{0,1\} \leftarrow \TPEvidenceVerify(\ParmsTP,r_{i\,t},x_{i\,t},\Evidence_t,t)$: run by each user. It
takes as input the system parameters $\ParmsTP$, slot secret $r_{i\,t}$, measurement $x_{i\,t}$ of user $i$ for
time $t$, and the evidence $\Evidence_t$, and outputs \textsc{True} or \textsc{False}. 

\end{itemize}
\vspace{-0.075cm}
Given $(\ParmsTP, k_r) := \TPInitialize(1^\kappa)$,  $(r_{i\,t})_{i\in[n]} :=
\TPSlotSecretGen(\ParmsTP,k_r)$, a set of valid measurements $(x_{i\,t})_{i\in[n]} \in \TPVSpace$ for time
slot $t$, the scheme is correct if  
$\TPEvidenceVerify((\ParmsTP,r_{j\,t},x_{j\,t},\TPEvidenceGen(\ParmsTP,k_r,(x_{i\,t})_{i\in[n]},t),t)=1$,
for any slot secret $r_{j\,t} \in (r_{i\,t})_{i\in[n]}$ and the corresponding  measurement $x_{j\,t} \in
  (x_{i\,t})_{i\in[n]}$.

\subsection{Security Properties} 
We define two properties of PPTP: \emph{transparency} and \emph{privacy}. For simplicity, we focus on a single
operational cycle.

\vspace{1mm}\noindent{\bf Transparency.} 
This property means that each honest user \mbox{$i \in [n]$} is able to verify that a bill $\badbill_i$ is
incorrect if \mbox{$\badbill_i > B^*_i$}, where $B^*_i$ is the worst-case bill defined earlier in the section.
Furthermore, the user can check if $\badbill_i$ contains an incorrect measurement $\bar{x}_{i\,t}$, i.e.,
$\bar{x}_{i\,t}> x_{i\,t}$ or $\bar{x}_{i\,t}> \delta_t$.  
Finally, a dishonest user cannot blame the retailer
for misbehavior if it has behaved honestly.

We formulate transparency through a two-stage game $\GameP^{\Consistency}_{\PPPT,\Adversary}$
running between a challenger and an adversary $\Adversary$. In the first stage, the adversary can run the
PPTP scheme herself with the secret key $k_r$ and measurements of her own choice,
and outputs a time slot $t^*$ as challenge request. The challenger then randomly samples a set of honest
measurements $\vec{x}_{t^*}=(x_{i\,t^*})_{i\in[n]} \rand \TPVSpace$ which are sent to $\Adversary$.  The goal
of $\Adversary$ is to generate a valid evidence $\bar{\Evidence}_{t^*}$ %
for a set of measurements $\bar{\vec{x}}_{t^*}=(\bar{x}_{i\,t^*})_{i\in[n]}$, such that
  $\bar{\vec{x}}_{t^*}\neq \vec{x}_{t^*}$, which can pass the verification for an honest user $j^*$'s
  measurement $x_{j^*,t^*} \in \vec{x}_{t^*}$, meaning that 
  $\TPEvidenceVerify(\ParmsTP,r_{j\,t^*},x_{j\,t^*},\bar{\Evidence}_{t^*},t^*)=1$.

\vspace{1mm}\noindent{\bf Privacy.} This property means that a user $i$ cannot learn
privacy-sensitive meter readings $x_{j\,t}$ of another user \mbox{$j \in [n] \backslash \{i\}$} for any $t \in
[k]$. %

We define a game $\GameP^{\Privacy}_{\PPPT,\Adversary}$ to formulate the privacy via
indistinguishability. The adversary $\Adversary$ chooses measurements $(x_{i\,t})_{i\in[n],t \in [k]}$ for all
users, and an honest user $j^*$ and time slot $t^*$ as the challenge in the game.  The challenger selects two
values $x^0_{j^*\,t^*}, x^0_{j^*\,t^*}$ that lead to the same bills for all users $i \neq j^*$. Specifically,
it sets $x^0_{j^*\,t^*}:=x_{j^*\,t^*}$, and $x^1_{j^*,t^*} \rand \TPVSpace$, such that
$x^0_{j^*,t^*} \neq x^1_{j^*,t^*}$ and  $(\bigwedge^{n}_{i=1,i\neq j^*} B^0_i = B^1_i)=1$, where $B^b_i =
p_{i\,t^*}(x_{i\,t^*},x^{b,*}_{t^*}) x_{i\,t^*}+ \sum_{t=1,t\neq t^*}^k p_{i\,t}(x_{i\,t},x^*_t) x_{i\,t}$
and $x^{b,*}_{t^*}=\min(x^{b}_{j^*,t^*},d_{t^*})+\sum_{i=1,i\neq j^*}^n \min(x_{i\,t^*},d_{t^*})$. Next, the
challenger generates the evidence for time $t^*$ based on $(x_{i\,t^*})_{i\in[n],i\neq
j^*}||x^b_{j^*\,t^*}$ and $b \rand \{0,1\}$. The goal of $\Adversary$ in this game is to
distinguish whether her measurement $x_{j^*\,t^*}$ is used to generate the challenge evidence
$\Evidence_{t^*}$.
Note that the constraint that the values chosen by the challenger leads to the same bill for users $i \neq
j^*$ is important.
Otherwise, the distinct bills %
would allow the adversary to trivially distinguish the bit
$b$.  During the game, the adversary can ask oracle $\OracleEGen$ to get the evidences for any measurements,
and oracle $\OracleCorrupt$ to reveal the slot secrets. But it cannot ask either $\OracleEGen(\cdot,
t^*)$ or $\OracleCorrupt(j^*,t^*)$. %

\begin{table}[!ht]
\vspace{-0.4cm}
\resizebox{\columnwidth}{!}{
\begin{tabular}{|ll|}
\hline 
$\underline{\GameP^{\Consistency}_{\PPPT,\Adversary}(\kappa)}$                                                                                                         & $\underline{\GameP^{\Privacy}_{\PPPT,\Adversary}(\kappa})$                                                               \\

$\Qlist:=\emptyset$                                                                                                                                            & $\Qlist:=\emptyset$; $\CorruptList :=\emptyset$; $b \rand \{0,1\}$                                               \\

$(\ParmsTP, k_r) \leftarrow \TPInitialize(1^\kappa)$                                                                                                           & $(\ParmsTP, k_r) \leftarrow \TPInitialize(1^\kappa)$                                                             \\

$(state,t^*)\leftarrow \Adversary(\ParmsTP,k_r)$                                                        & $(state,j^*,t^*, (x_{i\,t})_{i\in[n],t \in [k]}) \leftarrow \Adversary^{\OracleEGen(\cdot,\cdot),\OracleCorrupt(\cdot,\cdot)}(\ParmsTP)$          \\

$\vec{x}_{t^*}=(x_{i\,t^*})_{i\in[n]} \rand \TPVSpace$, s.t., $x_{i\,t^*} \leq \delta_t$ for $i \in [n]$                                                                                                                     &      $x^0_{j^*\,t^*}:=x_{j^*\,t^*}$                                      \\

  $(\bar{\Evidence}_{t^*},(\bar{x}_{i\,t^*})_{i\in[n]},j^*) \leftarrow \Adversary(\ParmsTP,k_r,state,(x_{i\,t^*})_{i\in[n]})$                                                                           & $x^1_{j^*\,t^*}\rand \TPVSpace$, s.t., $(x^0_{j^*\,t^*}\neq x^1_{j^*\,t^*})$ and $ (\bigwedge^{n}_{i=1,i\neq j^*} B^0_i = B^1_i)=1$ \\
  
 $\bar{\vec{x}}_{t^*}=(\bar{x}_{i\,t^*})_{i\in[n]}$ & $ \quad \text{where } B^b_i = p_{i\,t^*}(x_{i\,t^*},x^{b,*}_{t^*}) x_{i\,t^*}+ \sum_{t=1,t\neq t^*}^k p_{i\,t}(x_{i\,t},x^*_t) x_{i\,t} $  \\

Return $ (\vec{x}_{t^*}\neq \bar{\vec{x}}_{t^*}) \wedge (\bigvee^{n}_{i=1} \bar{x}_{i\,t^*}> \delta_t)$ &          $\quad \text{and } x^{b,*}_{t^*}=\min(x^{b}_{j^*,t^*},\delta_{t^*})+\sum_{i=1,i\neq j^*}^n \min(x_{i\,t^*},\delta_{t^*})$                  \\

 $\quad \wedge (\bar{\Evidence}_{t^*} = \TPEvidenceGen(\ParmsTP,k_r,(\bar{x}_{i\,t^*})_{i\in[n]},t^*)$                                                                     &     $\Evidence_{t^*} := \TPEvidenceGen(\ParmsTP,k_r,((x_{i\,t^*})_{i\in[n]\backslash j^*}||x^b_{j^*,t^*}),t^*)$    \\
 
 $\quad \wedge (\TPEvidenceVerify(\ParmsTP,r_{j^*\,t^*},x_{j^*\,t^*},\bar{\Evidence}_{t^*},t^*)=1)$                                                                    &            $b' \leftarrow \Adversary^{\OracleEGen(\cdot,\cdot),\OracleCorrupt(\cdot,\cdot)}(\ParmsTP,\Evidence_{t^*},state)$                                                                                                   \\

                                                            &   Return $(b=b') \wedge   (t^* \notin \Qlist)\wedge (r_{j^*,t^*} \notin \CorruptList)$      \\

                                                             & \\ 

$\underline{\OracleEGen((x_{i\,t})_{i\in[n]},t)}$                                                                                                              & $\underline{\OracleCorrupt(i,t)}$                                                                                \\
$E_t:=\TPEvidenceGen(\ParmsTP,k_r,(x_{i\,t})_{i\in[n]},t)$                                                                                                     & $(r_{i\,t})_{i\in[n]} :=  \TPSlotSecretGen(\ParmsTP,k_r,t)$                                                      \\

$(E_t,t,(x_{i\,t})_{i\in[n]}) \rightarrow \Qlist$                                                                                                              & $r_{i\,t}\rightarrow \CorruptList $                                                                              \\

Return $\Evidence_t$                                                                                                   & Return $r_{i\,t}$                                                                                                \\
                                                                                                                   \hline                                                                                                        
\end{tabular}
}
\vspace{-0.3cm}
\end{table}

\begin{definition}
A privacy-preserving transparent pricing scheme $\PPPT$ is secure if the advantages  $\Adv^{\Consistency}_{\PPPT,\Adversary}(\kappa)=\Pr[ \GameP^{\Consistency}_{\PPPT,\Adversary}=1]$ and  $\Adv^{\Privacy}_{\PPPT,\Adversary}(\kappa)=\left|\Pr[ \GameP^{\Privacy}_{\PPPT,\Adversary}(\kappa)=1] -1/2 \right|$ of any PPT adversaries $\Adversary$ in the corresponding games are negligible. 
\label{def:security}
\end{definition}

\section{Baseline Protocol}
\label{sec:baseline}

In this section, we present a simple instantiation of PPTP. This baseline protocol meets the security
definition above. In particular,  the retailer creates a commitment and zero-knowledge range proof for each
user's measurement, and shares the full set of commitments and proofs with all users. Each honest user verifies all the
steps of the protocol to detect misbehavior. This protocol incurs a large performance overhead, which we
address in next section with a more efficient protocol. In the following, we focus on  
the algorithms for a single time period $t \in [k]$, and discuss how to generalize it to a full operation
cycle. 

\subsection{Preliminaries}
We briefly review the syntax and security properties of the cryptographic primitives used in our protocol. 

\vspace{1mm}\noindent\textbf{Commitment Scheme.} A commitment scheme $\Comm$ consists of two algorithms. $
\CommSetup(1^\kappa)$ takes as input the security parameter $1^\kappa$, and outputs commitment
parameters $\ParmsC$. The algorithm $\CommCommit(\ParmsC, v,r)$ takes as input parameters $\ParmsC $, a value $v$, and
randomness $r$, and outputs a commitment $c$.  %
The scheme is \emph{hiding} when $c$ reveals nothing of the committed value, and \emph{binding} when the
commitment cannot be opened with two different random values. The scheme is \emph{collision-resistant} if for
any $(v_0, v_1) \in \VSpace$ and $(r_0,r_1) \in \RSpace$ such that $v_0 \neq v_1$, the probability that
$\CommCommit(\ParmsC, v_0,r_0) = \CommCommit(\ParmsC, v_1,r_1)$ is negligible. Finally, the scheme is
(additively) homomorphic if, for any \mbox{$(v_0, v_1) \in \VSpace$} and
$(r_0,r_1) \in \RSpace$, we have 
$$
\CommCommit(\ParmsC, v_0,r_0) + \CommCommit(\ParmsC, v_1,r_1) =
\CommCommit(\ParmsC, v_0+v_1,r_0+r_1).
$$ 

\vspace{1mm}\noindent\textbf{Non-interactive Zero-knowledge.} 
Let $\RelationSpace$ be an efficiently computable relation of the form  $(st, w) \in \RelationSpace$, where
$st$ is the statement and $w$ is the witness of $st$. A non-interactive proof system $\NIZK$ for
$\RelationSpace$ consists of three algorithms. $\NIZKSetup(1^\kappa)$ which takes as input the security
parameter $1^\kappa$ and outputs the system parameters $\ParmsZK$. $\NIZKProve(\ParmsZK, st,w)$ takes as input
system parameters $\ParmsZK$ and a pair $(st,w)$ and outputs a proof $\pi$. $\NIZKVerify(\ParmsZK, st,
\pi)$ takes as input $\ParmsZK$, a statement $st$, and a proof
$\pi$, and outputs \true $(1)$ or \false $(0)$.

The proof system $\NIZK$ satisfies {\em zero-knowledge} if the generated proofs reveal nothing regarding the
corresponding witnesses, and {\em simulation-extractability} if for any proof generated
by the adversary, there exists an efficient algorithm to extract the corresponding witnesses with a trap door.
In this work, we use $\NIZK$ over the family of relations $\RelationSpace =(\PPTPRelation_{\Vmax})_{\Vmax \in
\mathbb{N}}$, where $\Vmax$ can be either $\gamma_t$ or $\delta_t$, and  
\[
\PPTPRelation_{\Vmax}=\{(c,\Vmax), (v,r) \big| c=\CommCommit(\ParmsC, v, r) \wedge v \in [0, \Vmax] \}.
\]
\noindent\textbf{Pseudo-Random Functions.}
A pseudo-random function (PRF) family consists of two algorithms. 
$\PRFKGen(1^\kappa)$ is takes as input the security parameter $1^\kappa$ and outputs a random key
$k$. $\PRFEval(k,x)$ is takes as input the random key $k$ and a
message $x$, and outputs a pseudo-random value $r$.%

\subsection{Instantiation}

The four main algorithms of Section~\ref{sec:security_model} are instantiated as follows.
\vspace{-0.075cm}
\setlength{\leftmargini}{0.3em}
\begin{itemize}
\setlength\itemsep{0.2em}

\item[]$(\ParmsTP, k_r) \leftarrow \TPInitialize(1^\kappa)$: this algorithm generates the secret key $k_r:=
\PRFKGen(1^\kappa)$, and initializes the commitment and non-interactive zero-knowledge schemes as $\ParmsC :=
\CommSetup(1^\kappa)$ and $\ParmsZK :=\NIZKSetup(1^\kappa)$, respectively. The retailer publishes the system
parameters $\ParmsTP = (\ParmsC,\ParmsZK, \alpha_t, \beta_t, \gamma_t, \delta_t)$ and keeps the secret key
$k_r$ private.  

\item[] $(r_{i\,t})_{i\in[n]} \leftarrow  \TPSlotSecretGen(\ParmsTP,k_r,t)$: this algorithm uses the
retailer's secret key to generate the slot secrets as $r_{i\,t} := \PRFEval(k_r, i||t) $ for $i \in [n]$. Each
slot secret $r_{i\,t}$ is sent to the corresponding user $i$ via a secure channel and is stored privately by the retailer and user $i$. 

\item[] $\Evidence_t \leftarrow \TPEvidenceGen(\ParmsTP,(x_{i\,t}, r_{i\,t})_{i\in[n]}, t)$: this algorithm
computes for each user $c_{i\,t} := \CommCommit(\ParmsC, x_{i\,t}, r_{i\,t})$  and generates the proof
$\pi_{i\,t} := \NIZKProve(\ParmsZK, st, w)$ for the statement $st=(c_{i\,t},\delta_t)$ and witness $w =
(x_{i\,t},r_{i\,t})$.
The retailer computes the sum of the values $x_t^* = \sum_{i=1}^n x_{i\,t}$, the sum of the
random seeds $r_t^* = \sum_{i=1}^n r_{i\,t}$, $c_t^* := \CommCommit(\ParmsC, x_t^*, r_t^*)$, and a proof
$\pi_t^* := \NIZKProve(\ParmsZK, st, w)$ for $st=(c_t^*,\gamma_t)$ and witness $w = (x_t^*,r_t^*)$.  Finally,
the retailer shares $\Evidence_t = (c_t^*, \pi_t^*, (c_{i\,t}, \pi_{i\,t})_{i\in[n]})$ with each user,
and puts a hash $h_t = H(E_t)$ on the bulletin board.
     
\item[] $\{0,1\} \leftarrow \TPEvidenceVerify(\ParmsTP,r_{i\,t},x_{i\,t},\Evidence_t,t)$: this algorithm
executes four subroutines, each taking parts of $\Evidence_t = (c_t^*, \pi_t^*,
(c_{i\,t}, \pi_{i\,t})_{i\in[n]})$ as input, and returns \true if all four subroutines return 1. 

\setlength{\leftmargini}{0.3em}
\begin{itemize}
    \item[] $\{0, 1\} \leftarrow \mathsf{VerifyConsistency(}\Evidence_t, h\texttt{)}$: 
    computes $\bar{h} = H(\Evidence_t)$, and outputs \true if $\bar{h} = h$, and \false otherwise. 
    
    \item[] $\{0, 1\} \leftarrow \mathsf{VerifyCommitment(}\ParmsC,x_{i\,t}, c_{i\,t}, t\texttt{)}$: computes 
    $\bar{c}_{i\,t} := \CommCommit(\ParmsC, x_{i\,t}, r_{i\,t})$ and outputs \true if $\bar{c}_{i\,t} =
    c_{i\,t}$, and \false otherwise. 
    
    \item[] $\{0, 1\} \leftarrow \mathsf{VerifySum(}\ParmsZK,c_t^*, \pi_t^*, (c_{i\,t})_{i\in[n]}\texttt{)}$:
    computes $\bar{c}_t^* = \sum_{i=1}^n c_{i\,t}$. It checks if  
        $\bar{c}^* = c^*$. It then checks that $c^*$ and $\pi_t^*$ match, and executes $\NIZKVerify(\ParmsZK,
        st, \pi_t^*)$ for the statement $st=(c^*,\gamma_t)$. The functions outputs \true if all three
        checks are successful, and \false otherwise. 
    
    \item[] $\{0, 1\} \leftarrow \mathsf{VerifyRangeProofs(}\ParmsZK,(c_{i\,t}, \pi_{i\,t})_{i \in [n]}\texttt{)}$:
    checks that for each $i$, the commitment $c_{i\,t}$ and the proof $\pi_t$ match, then
    executes $\NIZKVerify(\ParmsZK, st, \pi_{i\,t})$ for the statement $st=(c_{i\,t},\delta_t)$. It  
    outputs \true if all proofs are valid and match the commitment, and \false otherwise.
\end{itemize}

\end{itemize}

\vspace{1mm}{\noindent \bf Operational cycles.}
For a full operational cycle, the protocol above is executed for every time period $t \in [k]$.
The retailer computes $B_i$ and sends it to each user $i \in [n]$, who then verifies the bill since she knows $x_{i\,t}$ and
$x^*_t$ for each time period.

\subsection{Security Analysis} 
The baseline protocol described in this section is secure according to Definition~\ref{def:security}. The proofs for the
following two theorems are included in the Appendix~\ref{sec:security_proofs}.  

\begin{theorem}\label{thm:pptp_transparancy_baseline}
If the commitment scheme $\Comm$ is additively homomorphic and satisfies binding property, and non-interactive
zero-knowledge proof system $\NIZK$ is simulation-extractable, then the baseline protocol provides transparency.
\end{theorem}

\begin{theorem}\label{thm:pptp_anonymity_baseline}
If the commitment scheme $\Comm$ is additively homomorphic and satisfies hiding property, the non-interactive
zero-knowledge proof system $\NIZK$ provides zero-knowledge, and pseudo-random function family $\PRF$ is
secure, then the baseline protocol satisfies privacy.  \end{theorem}

\subsection{Performance Analysis}
We analyze the computation cost of the protocol in terms of the number of invocations to the cryptographic
functions, and the bandwidth cost in terms of the sizes of the messages sent
over the network. We ignore the initiation cost and the cost of users sending measurements to the retailer, as these are independent of the implementation.  
In the following, $M_{c}$, $M_{\pi}$, $M_h$ respectively denote the message size of the commitments, proofs, and hashes.\footnote{Some range proof techniques, e.g., Bulletproofs \cite{bunz2018bulletproofs}, allow for
the aggregation of multiple proofs over the same range, leading to reduced bandwidth costs.}

The total costs per time slot for each entity are summarized in Table~\ref{tab:baseline_cost}. 
The server has to execute the $\CommCommit$ and $\NIZKProve$ $n+1$ times in the $\TPEvidenceGen$ algorithm:
$n$ times for each client's measurement $x_{i\,t}$ and once for the sum $x^*_t$. The $n+1$ commitments and
proofs are sent to each of the $n$ clients, meaning that a message of size $(n+1)(M_{c}+M_{\pi})$ is sent  
$n$ times. The server also computes $H(\Evidence_t)$, and sends this hash to the bulletin
board. Each client receives a single message of size $n(M_{c} + M_{\pi})$, and downloads the hash from the
bulletin board for verification. The client then 1) verifies the commitment of its own measurement by
executing $\CommCommit$ and checking whether its output matches the value sent by the retailer, 2) executes
$\NIZKVerify$ on each of the proofs, and 3) check whether the hash of the result matches the one on the
bulletin board. The load on the bulletin board consists of receiving a single hash from the retailer, and
sending it to the $n$ users.

\begin{table}[!ht]
\centering
\begin{scriptsize}
\caption{Performance cost of the baseline solution.
}
\makebox[\textwidth][c]{
\begin{tabular}{c|ccc|c}
\multicolumn{1}{c}{\multirow{2}{*}{Entity}} & \multicolumn{3}{c}{Computation} &
\multicolumn{1}{c}{\multirow{2}{*}{Bandwidth}} \\ 
& $\CommCommit$ & $\NIZKProve$ & $\NIZKVerify$ & \\ \hline
Server & $n+1$ & $n+1$ & $0$ & $n(n+1)(M_c + M_\pi) + M_h$ \\ \hline
Client & $1$ & $0$ & $n+1$ & $(n+1)(M_c+M_\pi) + M_h$ \\ \hline
Bulletin Board & $0$ & $0$ & $0$ & $(n+1)M_h$ \\ \hline
\end{tabular}
}
\label{tab:baseline_cost}
\setlength{\tabcolsep}{3pt}
\end{scriptsize}
\end{table}

\ignore{
\setlength{\tabcolsep}{3pt}
\begin{table}[!ht]
    \centering
    \caption{Computation and bandwidth costs of the three main entities in the baseline solution.}
    \begin{tabular}{c|cccc|ccc}
        \multicolumn{1}{c}{} & \multicolumn{4}{c}{processing} & \multicolumn{3}{c}{bandwidth} \\
        \multicolumn{1}{c}{entity} & $\CommCommit$ & $\NIZKProve$ & $\NIZKVerify$ & \multicolumn{1}{c}{$H\;$} &  $M_c$ & $M_{\pi}$ & $M_h$ \\ \toprule
        server & $\Theta(n)$ & $\Theta(n)$ & 0 & $\Theta(1)$ & $\Theta(n^2)$ & $\Theta(n^2)$ & $\Theta(1)$ \\
        client & $\Theta(1)$ & 0 & $\Theta(n)$ & $\Theta(1)$ & $\Theta(n)$ & $\Theta(n)$ & $\Theta(1)$ \\
        bulletin board$\;$ & 0 & 0 & 0 & 0 & 0 & 0 & $\Theta(n)$ \\ \bottomrule
    \end{tabular}
    \label{tab:baseline_cost}
\end{table}
}

\subsection{Discussion}
\label{sec:baseline_discussion}
We consider the computation cost at the server, which increases linearly in $n$, to be acceptable
because the server is often equipped with powerful CPUs, and the operations are parallelizable. The bandwidth
cost increases quadratically, but the server can offload the messages to a cloud storage or content delivery network that can
distribute the messages in a scalable and cost-effective fashion. We discuss a blockchain-based extension to the
baseline protocol that reduces the costs at the client, which is important since the clients are likely
running on edge devices close to the meters, or even running directly on the meter firmware.

\begin{table}[!ht]
\centering
\begin{scriptsize}
\caption{Performance cost of the blockchain solution. $\mathsf{BC}$ represents the processing of read requests
at the blockchain smart contract. $M_{bc}$ is the message containing the blockhain proof that a commitment is stored in the chain. 
}
\makebox[\textwidth][c]{
\begin{tabular}{c|cccc|c}
\multicolumn{1}{c}{\multirow{2}{*}{Entity}} & \multicolumn{4}{c}{Computation} & \multicolumn{1}{c}{\multirow{2}{*}{Bandwidth}} \\ 
& $\CommCommit$ & $\NIZKProve$ & $\NIZKVerify$ & $\mathsf{BC}$ & \\ \hline
Server & $n+1$ & $n+1$ & $0$ & $0$ & $(n+1)(M_c + M_\pi)$ \\ \hline
Client & $1$ & $0$ & $0$ & $1$ & $M_c + M_{bc}$ \\ \hline
Bulletin Board & $0$ & $0$ & $n+1$ & $n + 1$ & $(2n+1)M_{c} + nM_{bc} + (n+1)M_\pi$ \\ \hline
\end{tabular}
}
\label{tab:baseline_imprv_cost}
\setlength{\tabcolsep}{3pt}
\end{scriptsize}
\end{table}

We note that for verification, only the $\mathsf{VerifyCommitment}$ algorithm requires the private input. In
other words, most of the verification can be done publicly by another party without compromising client privacy.
The $\mathsf{VerifyRangeProof}$ algorithm, which is the most expensive, can therefore be performed by a trusted third
party with more resources than the client.  As a first improvement over the baseline solution, we propose to leverage a blockchain as the trusted party, and
implement the range proof verification in a \textit{smart contract}. 

The server sends the proofs and commitments to a blockchain smart contract, which executes
$\mathsf{VerifyConsistency}$, $\mathsf{VerifySum}$, and $\mathsf{VerifyRangeProofs}$. If the verifications
pass, the blockchain stores the messages. The client $i$ only computes the commitment $c_{i\,t}$ using its own measurement $x_{i\,t}$, and verifies that it was
included in the blockchain by reading blockchain storage directly without incurring consensus
overhead. The client does not need to run $\NIZKVerify$ on any of the other users' range proofs, but the trade-off here is that the blockchain incurs significant
overhead: generating $n$ proofs for read requests from the user, and executing the smart contract for
verification. We summarize the cost for each entity in Table~\ref{tab:baseline_imprv_cost} in the Appendix. 

\section{Merkle Tree Protocol}
\label{sec:solution}
As discussed in Section~\ref{sec:baseline_discussion}, the baseline solution of Section~\ref{sec:baseline} can
be improved  by offloading all range proof verification to a trusted third party such as the bulletin board.  However,
this puts a large computational burden on the bulletin board -- far beyond the capacity of, for example, the
Ethereum blockchain.
In this section, we take inspiration from Certificate Transparency \cite{CCS:ChaseM16} and describe another PPTP instantiation based on a modified Merkle tree. The protocol reduces the
 communication and computation costs at the user, and therefore is more practical.  
 In each period the retailer sends to each user a proof of the inclusion and a range proof for the sum $x^*_t$, which due to the homomorphism of our commitment scheme corresponds to the value in the root of the tree.

\subsection{Overview}
\begin{figure}
\vspace{-0.4cm}
\centering
	\includegraphics[width=0.7\linewidth]{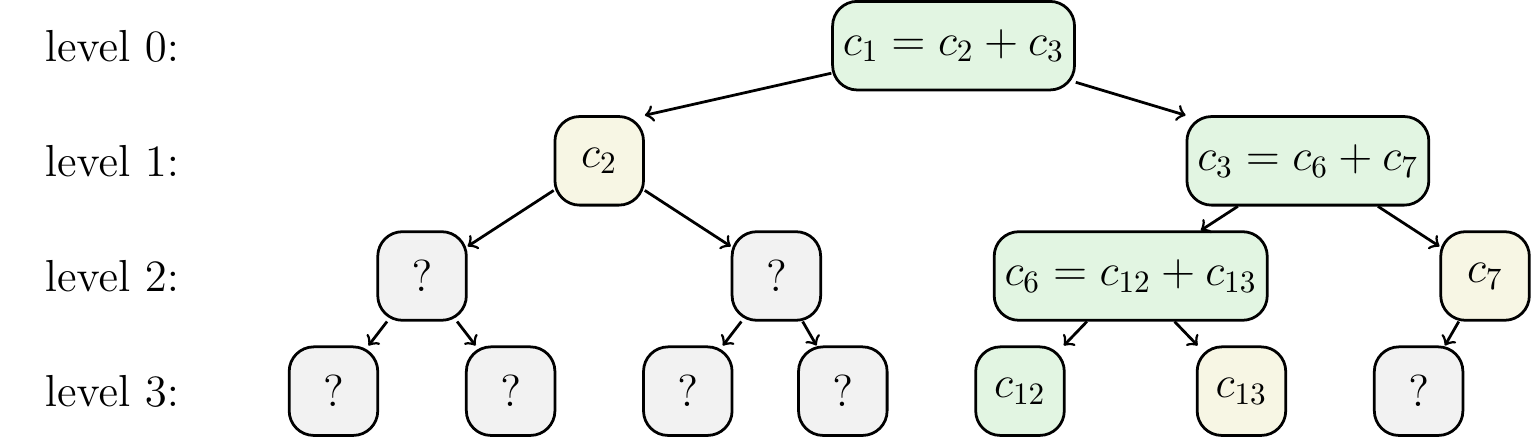}
	\caption{Example of a modified Merkle tree with 7 leaves.}
	\label{fig:commitmenttree}
	\vspace{-0.6cm}
\end{figure}

Merkle trees are an extension of \emph{binary trees}. 
For any node $\node$, let $\textsc{left}(\node)$ be its first child and $\textsc{right}(\node)$ its second
child. If node $\node$ has no second child, then $\textsc{right}(\node) = 0$, and if it also has no first
child then $\textsc{left}(\node) = 0$.
In the original Merkle tree, each node $\node$ contains a hash value $h_{\node}$ of its children.  We modify
it by storing at each node $\node$ a commitment $c_{\node}$.
For each user $i \in [n]$ we set the commitment in the $i$th leaf equal to $c_{i} = \CommCommit(\ParmsC, x_{i\,t}, r_{i\,t})$. The commitments in the other nodes are computed as: \begin{equation}
\begin{array}{c} 
	c_{\node} = c_{\textsc{left}(\node)} + c_{\textsc{right}(\node)},
\end{array}
\label{eq:commitmenttree}
\end{equation}
such that and $c_0 = \CommCommit(\ParmsC, 0, 0)$. The inclusion proof for node $i$ includes the
root, $i$'s leaf, the intermediate nodes that lie on the path between them, and the children of these nodes.
Figure~\ref{fig:commitmenttree} shows an example, where the green nodes represent the path between the
leaf and the root and the yellow nodes the children of those nodes.

\vspace{1mm}{\noindent \bf Assumptions.} In this section, we describe a protocol that is secure under the
following assumptions. There are a number of powerful entities, called {\em auditors}, that
perform verification on behalf of the users. In the following, we assume that auditors do not share measurements themselves (although in practice a single entity may act as a user and an auditor). At least one auditor is honest, and at most $f$ are dishonest.  
Furthermore,
the maximum time for an auditor to finish its verification and have its messages stored on the bulletin
board, $T$, is known. 
We discuss other protocols that do not rely on auditors in the
Appendix~\ref{sec:no_auditor}. 

\subsection{Instantiation}

The four algorithms of Section~\ref{sec:security_model} are instantiated as follows.
\vspace{0.075cm}

\setlength{\leftmargini}{0.3em}
\begin{itemize}
\setlength\itemsep{0.5em}

\item[]$\TPInitialize,\ \TPSlotSecretGen$: they are the same as in Section~\ref{sec:baseline}.  

\item[] $\Evidence_t \leftarrow \TPEvidenceGen(\ParmsTP,(x_{i\,t}, r_{i\,t})_{i\in[n]}, t)$: this algorithm
constructs a Merkle tree described above, using the measurements and slot randomness as input.
For each user, the evidence $E_t$ contains the inclusion proof $G$ and a zero-knowledge proof $\pi_t^*$ that the sum $x_t^*$ (whose commitment $c_t^*$ is the root of $G$) is within the range
$[0, \gamma_t]$. For auditors, the evidence also includes the commitments and range proofs $(c_{i\,t}, \pi_{i\,t})$ for all leaf nodes $i \in [n]$.
The retailer also uploads the commitment in the root node to the
public bulletin board as $c_{h}$.

\item[] $\{0,1\} \leftarrow \TPEvidenceVerify(\ParmsTP,r_{i\,t},x_{i\,t},\Evidence_t,t)$: 
this algorithm consists of two phases. In the first phase, it execute the following subroutines.
\vspace{0.2cm}
\setlength\itemsep{0.5em}
\begin{itemize}
    \item[] $\{0, 1\} \leftarrow \mathsf{VerifyConsistency(}G, c_{h}\texttt{)}$: verifies that the value of $c_{h}$ on the bulletin board matches with the root of the  
    inclusion proof $G$. %
    
    \item[] $\{0, 1\} \leftarrow \mathsf{VerifyCommitment(}\ParmsC,r_{i\,t},x_{i\,t},G\texttt{)}$: 
    if the user is an auditor, returns \true. Else, it computes $\bar{c}_{i\,t} := \CommCommit(\ParmsC,
      x_{i\,t}, r_{i\,t})$, and verifies that $\bar{c}_{i\,t} = c_{\nodea}$, where $\nodea$ is the leaf node
      in the user's inclusion proof $G$. 

    \item[] $\{0, 1\} \leftarrow \mathsf{VerifyInclusionProof(}G\texttt{)}$:
    if the user is an auditor, returns \true. Else, it verifies that the inclusion proof is valid, i.e., that the commitment $c$ in each
      intermediate core node $\node \in G$ indeed follows from applying \eqref{eq:commitmenttree} to the commitments in
      $\node$'s children.

    \item[] $\{0, 1\} \leftarrow \mathsf{VerifySum(}\ParmsZK,G,\pi_t^*\texttt{)}$:
    if the user is an auditor, returns \true. Else, it verifies that the proof $\pi_t^*$ matches the commitment $c_t^*$ in the root of $G$, and executes $\NIZKVerify(\ParmsZK,
        st, \pi_t^*)$ for the statement $st=(c_t^*,\gamma_t)$. 

    \item[] $\{0, 1\} \leftarrow \mathsf{VerifyTree}(\ParmsZK, (c_{i\,t}, \pi_{i\,t})_{i\in[n]})$: if the user is not an auditor, returns
    \true. Else, it executes $\mathsf{VerifyRangeProofs}(\ParmsZK, (c_{i\,t}, \pi_{i\,t})_{i\in[n]})$ from Section~\ref{sec:baseline} for each user $i \in [n]$. Then, it
    checks whether the non-leaf nodes of the tree are correct, i.e., whether they follow from applying \eqref{eq:commitmenttree}.
    It returns \true if all checks are successful, and \false otherwise. 
  \end{itemize}
  \vspace{1mm}
  In the second phase, if the user is an auditor, it signs and publishes a message to the bulletin board: a failed range proof for a
  leaf node if it exists, the value $1$ if all four subroutines in the fist phase succeed, or an empty message
  otherwise. The algorithm then returns true or false depending on the result of the first phase. The
  non-auditor user checks if there is a proof of incorrect range in the bulletin board, if yes, it verifies
  the proof and returns false. Otherwise, it waits until there are $f+1$ signed empty messages on the bulletin
  and returns \true, or returns \false after time $T$. 
\end{itemize}

\subsection{Security Analysis} 
We show that this protocol is a secure PPTP, by proving the following theorems. The proofs are included in the
Appendix. 

\begin{theorem}\label{thm:pol_transparency_merkle}
If the commitment scheme $\Comm$ is additively homomorphic and satisfies binding property, and non-interactive
zero-knowledge proof system $\NIZK$ is simulation-extractable, and the assumptions about $f$ and $T$ hold,
then the Merkle tree protocol provides transparency. 

\end{theorem}

\begin{theorem}\label{thm:pol_anonymity_merkle}
With the same assumptions in Theorem~\ref{thm:pptp_anonymity_baseline}, the Merkle tree  protocol satisfies privacy. 
\end{theorem}

\subsection{Performance Analysis}
We analyze the performance of the Merkle tree protocol using the same metrics as in the analysis of the
baseline protocol. For simplicty, we assume $n = 2^m$ for some $m \in \mathbb{N}$, so that the total number of
nodes in the tree equals $2^{m+1} - 1 = 2n - 1$. The retailer generates commitments for all $2n-1$ nodes and generates range proofs for the root and the $n$ leaf nodes. It only sends the commitments for the nodes in the  
inclusion proof -- this consists of $m = \log_{2} n$ nodes on the path and $m + 1$ children, so $2m+1$ in total. 
Table~\ref{tab:merkle_cost} summarizes the cost.  

We can also extend the protocol to use a blockchain as described in  Section~\ref{sec:baseline_discussion}. In
particular, we can offload the range proofs to the smart contracts, hence making the blockchain an auditor. 

\begin{table}[!ht]
\centering
\begin{scriptsize}
\caption{Performance cost of the Merkle tree solution, assuming $m = \log_2(n)$ and $f+1$ auditors who each report to the bulletin board. Note that a user can simultaneously act as a normal client and as an auditor.}\makebox[\textwidth][c]{
\begin{tabular}{c|ccc|c}
\multicolumn{1}{c}{\multirow{2}{*}{Entity}} & \multicolumn{3}{c}{Computation} & \multicolumn{1}{c}{\multirow{2}{*}{Bandwidth}} \\ 
& $\CommCommit$ & $\NIZKProve$ & $\NIZKVerify$ & \\ \hline
Server & $2n-1$ & $n+1$ & $0$ & $(n+1) (f+1) (M_c + M_{\pi}) + n(2m+1)M_c$ \\ \hline %
Normal Client & $1$ & $0$ & $1$ & $(2m + 1)M_c + (f+1) M_{bc}$ \\ \hline
Auditor & $0$ & $0$ & $n$ & $(n+1) (M_c + M_{\pi}) + M_{bc}$ \\ \hline
Bulletin Board & $0$ & $0$ & $0$ & $(n+1)(f+1) M_{bc}$ \\ \hline %
\end{tabular}
}
\label{tab:merkle_cost}
\setlength{\tabcolsep}{3pt}
\end{scriptsize}
\end{table}

\section{Implementation}
\label{sec:implementation}
We implement both the baseline and Merkle tree protocol in Go. To do so, we extend the zero-knowledge range proof
library by Morais et al.\ \cite{morais2019survey}.\footnote{\url{https://github.com/ing-bank/zkrp}} In  
particular, we use their implementation of Bulletproofs \cite{bunz2018bulletproofs}  which uses Pedersen
commitment with the \texttt{secp256k1} elliptic curve \cite{brown2010standards} as a commitment scheme. We
parallelize proof generations and verification by exploiting Go's multi-threading. We release the source code for our experiments 
at {\url{https://github.com/aungmawjj/zkrp}}.  

We evaluate the performance of our protocols in terms of the computation and bandwidth cost. We also examine
how the performance scales with more hardware resources. There are no baselines from the related literature in our evaluation, because we
are unaware of other protocols that address the same problem and have the same security guarantee as ours. We
run the retailer's server on AWS EC2 with varying numbers of cores. We run the client on a Raspberry Pi (RPI)
3 with 4 CPUs and 1GB RAM. 

\begin{figure}
\vspace{-0.5cm}
\centering
    \begin{subfigure}[t]{.24\textwidth}
      \centering
      \includegraphics[width=\linewidth]{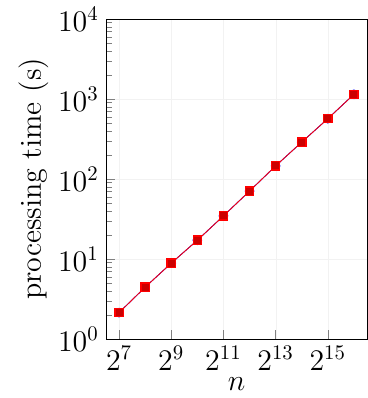}\vspace{-0.15cm}  
      \caption{}
      \label{fig:results-first}
    \end{subfigure}
    \begin{subfigure}[t]{.24\textwidth}
      \centering
      \includegraphics[width=\linewidth]{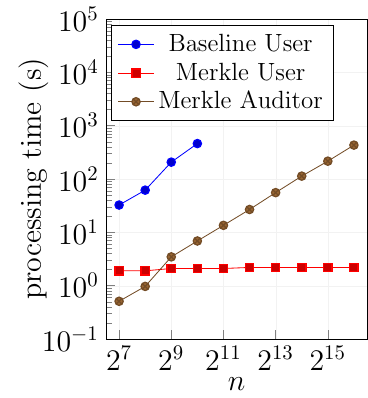}\vspace{-0.15cm}  
      \caption{}
      \label{fig:results-second}
    \end{subfigure}
    \begin{subfigure}[t]{.24\textwidth}
      \centering
      \includegraphics[width=\linewidth]{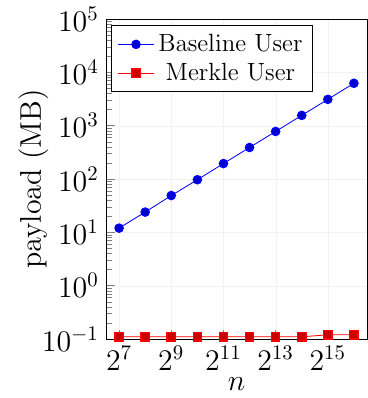}\vspace{-0.15cm}   
      \caption{}
      \label{fig:results-third}
    \end{subfigure}
    \begin{subfigure}[t]{.24\textwidth}
      \centering
      \includegraphics[width=\linewidth]{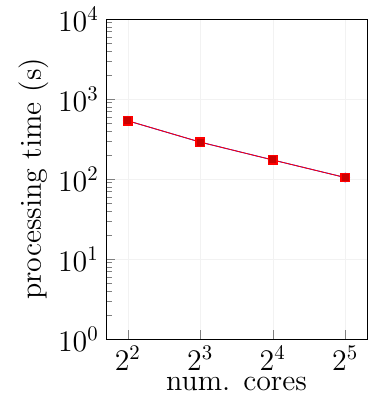}\vspace{-0.15cm}   
      \caption{}
      \label{fig:results-fourth}
    \end{subfigure}
    \vspace{-0.2cm}
    \caption{(a) Proof generation times, (b) proof verification times, (c) network cost. Experiments taking longer than $20$
    minutes are not included. (d) Proof generation times for $n = 2^{14}$.} \label{fig:results}
    \vspace{-0.5cm}
\end{figure}

{\noindent \bf Results.}
Figure~\ref{fig:results-first}, \ref{fig:results-second}, and \ref{fig:results-third} compare the costs of the two
protocols, with varying numbers of users, using a VM with 8 cores for the server. Proof generation takes just as long for both protocols, so the lines overlap -- in particular, for $n=16\,384$,
it takes $289s$ for the two protocols to generate the proofs. In Figure~\ref{fig:results-second}, the auditor runs on the same VM as the server.
The verification time and communication cost at the client increase linearly for the baseline, but only
logarithmically for the Merkle tree protocol. Even with $n =1\,024$, verification takes $460s$ for
the baseline, which is significant considering that it must be done $k$ times per operational cycle. In
contrast, the Merkle tree protocol takes $2.1s$.

Figure~\ref{fig:results-fourth} shows that both protocols scale well, in terms of the proof generation cost,
with more CPU cores. With $n=16\,384$, the time for Merkle tree solution reduces from $492s$ with 8 cores, to
only $167s$ with 32 cores.

We remark that the client cost is practical, for example with $n=65\,536$ it takes $2.2s$ to verify a single
proof of roughly $110$kB and check $33$ commitments. However, this cost increases linearly with $k$. As a large value of $k$ implies fine-grained
pricing, there is a trade-off between performance and how fine-grained the pricing scheme can be.

\section{Related Work}
\label{sec:related}
Shi et al.\ \cite{shi2011privacy} propose a privacy-preserving protocol that can be applied to smart meters. 
As mentioned in the introduction, their threat model is different from ours, as we allow the retailer to know individual users' measurements, but not to distort them. 
Another difference is that their method is unable to decrypt if a single meter is offline.
In our setting, the retailer can set the measurements of offline users to any value that does not exceed $\delta_t$, but offline users otherwise have no impact. %
\'Acs and Castelluccia \cite{acs2011have} propose a related scheme for smart meters that is able to decrypt when some meters fail, although the protocol requires additional steps to do so. 
Erkin and Tsudik \cite{erkin2012private} consider a setting for sharing meter data that is entirely peer-to-peer, i.e., without a retailer/aggregator. 

Zhang et al.\ \cite{zhang2015integridb} propose \textsc{IntegriDB}, which allows data owners to outsource SQL queries on their databases in a secure way. Their approach -- which uses sorted Merkle trees -- can be used to store sums and perform range proofs on measurement data, but they use a different system and threat model than the one in Section~\ref{sec:security_model}. In particular, range proofs in their setting require that the values in the leaves of the tree are sorted, but whereas their setting has an entity that checks the structure of the tree (the `data owner'), there is no such entity in our model as the auditors do not know the leaves' underlying measurements.

Chase and Meiklejohn~\cite{CCS:ChaseM16} define the first formal security model for provable transparency. Our formalization of transparency is adapted from their model but with additional constraints on electricity measurements and bills. Unlike Chase and Meiklejohn's model, we particularly formulate the privacy definition in conjunction with transparency. Nevertheless, we leverage auditors as in~\cite{CCS:ChaseM16} to facilitate the proof verification in the Merkle tree scheme.

\section{Conclusions}
\label{sec:conclusions}
We have introduced Privacy-Preserving Transparency Pricing (PPTP) schemes, which incentivize consumers to schedule
their electricity in a way that reduces the costs for both the retailer and the users. We showed that our scheme is
secure against retailer misbehavior while preserving user privacy. 
 Our scheme can support tens of thousands of
users while remaining computationally feasible on low-capacity devices. 
In our scheme, the retailer is only able to increase its profits to the extent allowed by the number of adversarial users, which is limited. That is, the retailer can inflate demand to a degree determined by the fraction of adversarial users and the maximum value for readings $\delta_t$. This allows the retailer to occasionally charge higher prices, but only if the system-level threshold $\gamma_t$ would not be exceeded if the adversarial users were to report their true usage, but exceeded if the adversarial users report the maximum usage $\delta_t$.

Our future work includes improving the
protocol further to support even larger numbers of users. 
We note that our approach can be extended to other situations in which it is beneficial for users to pay fees based on
network-level demand -- e.g., water or gas distribution networks. We also believe that our method generalizes
to systems with multiple retailers, but we leave this as future work.
Another interesting direction for future work is to investigate how quickly user behavior settles into a stable efficient pattern given the dynamic pricing scheme presented in Section~\ref{sec:system_model}. We believe that this will mainly depend on 1) the degree to which user behavior fluctuates between cycles, and 2) the exact strategy that users use to move controllable loads from high-demand to low-demand periods.
Regarding the former, we know from \cite{mohsenian2010optimal} that the demand in a day can be predicted with high accuracy using the demand in the previous days. The latter is a game theory question, but we expect that this will be a partially random strategy: e.g., if the demand in some period exceeds the threshold due to a change in usage patterns, then each user shifts each controllable load in the high-demand periods away to a low-demand period with some probability $p$.

\section*{Acknowledgments}
This research/project is supported by the National Research Foundation (NRF), Prime Minister's Office, Singapore, under its National Cybersecurity R\&D Programme and administered by the National Satellite of Excellence in Design Science and Technology for Secure Critical Infrastructure, Award No. NSoE DeST-SCI2019-0009. This research is also supported by A*STAR under its RIE2020 Advanced Manufacturing and Engineering (AME) Industry Alignment Fund - Pre Positioning (IAF-PP) Award  A19D6a0053. Any opinions, findings and conclusions or recommendations expressed in this material are those of the author(s) and do not reflect the views of A*STAR. Finally, we thank the anonymous reviewers whose comments helped improve the paper.

\bibliographystyle{abbrv}
\bibliography{ref}
\vspace{-0.25cm}
\appendix
\label{sec:appendix}
\subsection*{Appendix A: A Protocol Without Auditors}
\vspace{-0.1cm}
\label{sec:no_auditor}

We describe a protocol that relies on users randomly checking each other's proofs. The only difference to the
protocol presented in Section~\ref{sec:solution} is the $\TPEvidenceVerify$ algorithm. 
In particular, the $\TPEvidenceVerify(\ParmsTP,r_{i\,t},x_{i\,t},\Evidence_t,t)$ algorithm consists of two
phases. In the first phase, it executes the subroutines of $\TPEvidenceVerify$ from 
Section~\ref{sec:solution} where the user is not an auditor. 
\ignore{
  If any of the subroutines returns false,
the algorithm returns false. 

\setlength\itemsep{0.5em}
\begin{itemize}
    \item[] $\{0, 1\} \leftarrow \mathsf{VerifyConsistency(}x^*, r^*, (c_{i\,t}, \pi_{i\,t})_{i\in[n]},
    h\texttt{)}$: verifies that the values of $c$ and $l$ on the bulletin board matches with the root of the  
    inclusion proof. %
    
    \item[] $\{0, 1\} \leftarrow \mathsf{VerifyCommitment(}x_{i\,t}, c_{i\,t}, t\texttt{)}$: 
    computes $\bar{c}_{i\,t} := \CommCommit(\ParmsC, x_{i\,t}, r_{i\,t})$, and verifies that $\bar{c}_{i\,t}
    = c_{\nodea}$, where $\nodea$ is the leaf node in the user's inclusion proof. 

    \item[] $\{0, 1\} \leftarrow \mathsf{VerifyInclusionProof(}x^*,r^*, (c_{i\,t})_{i\in[n]}\texttt{)}$:
    verifies that the values for $c$ and $l$ in each intermediate core node $\node$ indeed follow from
    applying \eqref{eq:commitmenttree} to the values in $\node$'s children.

    \item[] $\{0, 1\} \leftarrow \mathsf{VerifyRangeProofs(}c_{i\,t}, \pi_{i\,t}\texttt{)}$: 
    verifies that the commitment $\pi$ matches the commitment $c$ in each core and edge node, and validates the
    zero-knowledge proof $\pi$ in each core and edge node.
  \end{itemize}
}
In the second phase, the user randomly picks $z$ other users, and then requests the inclusion proof $G_j$ and runs $\mathsf{VerifyRangeProofs}(\ParmsZK, c_{j\,t}, \pi_{j\,t})$ and
$\mathsf{VerifyInclusionProof}(G_j)$ for every user $j$ that it picks. If the range proof verification fails for
$j$, the user publishes the invalid inclusion proof of $j$ to the bulletin board. If the verification succeeds for all
$z$ users, then the user waits for a certain time $T$ and checks if the bulletin board contains any incorrect range proof. The algorithm returns true if there is no such proof, and false otherwise.

\begin{theorem}\label{thm:pol_transparency_merkle_random}
Let $h$ be the number of honest users who each check $z$ proofs of other users, and let $f$ be the number of malicious users whose committed value exceeds $\delta_t$, such that $h+f \leq n$. Let $T$ be a bound on the amount of time before any evidence of misbehavior appears on the bulletin board.
If the commitment scheme $\Comm$ is additively homomorphic and satisfies the binding property, and if the non-interactive
zero-knowledge proof system $\NIZK$ is simulation-extractable, then the above protocol achieves
 transparency for sufficiently large $h$ or $z$. %
 \end{theorem}
 
 \vspace{1mm}\noindent\textbf{Proof of Theorem~\ref{thm:pol_transparency_merkle_random}.} 
In the following, we assume that in each period $t$, each user $i$ draws the same number of other users $z$ from $[n] \backslash \{i\}$ without replacement (as there is no need to check the same proof twice). The number $U_i$ of incorrect leaves that are drawn by $i$ is then a random variable with the \textit{hypergeometric distribution}, i.e.,
$$
\mathbb{P}(U_i = u) = \comb{f}{u}\comb{n-f-1}{z-u}/\comb{n-1}{z}.
$$
By substituting $u=0$, the probability that user $i$ draws $0$ incorrect leaves can therefore be shown to be equal to $\frac{(n-f-1)!}{(n-f-z-1)!} / \frac{(n-1)!}{(n-z-1)!} = \prod_{i=0}^{z-1} \frac{n-f-i-1}{n-i-1}$ which is bounded by $\left(\frac{n-f-1}{n-1}\right)^z$ because the highest element in the product occurs at $i=0$. If $h$ honest nodes perform this experiment, then the probability that none of them detect an error is therefore at most $\left(\frac{n-f-1}{n-1}\right)^{hz}$ because the honest nodes perform their experiment independently. This is therefore an upper bound on the probability of failure, i.e., not detecting misbehavior, and the bound vanishes if $h\rightarrow \infty$ or $z \rightarrow \infty$.

\vspace{-0.25cm}
\subsection*{Appendix B: Security Proofs}\label{sec:security_proofs}
\vspace{-0.1cm}

In the following, we present the proofs of the corresponding theorems.

\vspace{1mm}\noindent\textbf{Proof of Theorem~\ref{thm:pptp_transparancy_baseline}.} 
We give a sketch of proof here due to shortness of space. Recall that in the transparency game the adversary has to generate at least a dishonest measurement such that $\bar{x}_{i\,t^*}> d_t$. First, we claim that the commitment $C^*_{i\,t^*} =\CommCommit(\ParmsC, \bar{x}_{i\,t^*}, r_{i\,t^*})$ of $\bar{x}$  has a negligible probability to collide with the commitment $C_{i\,t^*} =\CommCommit(\ParmsC, x_{i\,t^*}, r_{i\,t^*})$ of the honest (challenge) measurement $x_{i\,t^*}$, i.e., $ C^*_{i\,t^*}= C_{i\,t^*}$. Since if such a collision occurs with non-negligible probability, we could build an efficient algorithm $\mathcal{B}$ which runs the transparency-adversary $\Adversary$ as a subroutine to break the binding security property of $\Comm$. 
Therefore, there is no ambiguity about the committed measurements after each user’s confirmation of the corresponding commitment.  Next, we show that no adversary $\Adversary$ can produce a proof $\pi^*_{i\,t^*}$ for a false statement $\bar{x}_{i\,t^*}> d_t$ which may lead to a false bill $B^*_{j^*}> B_j$ where $B_j$ is the bill determined by the challenge measurements. Note that the condition $\bar{x}_{i\,t^*}> d_t$ implies that it belongs to a relation which does not belong to the honest relation $\PPTPRelation$ for $x_{i\,t^*}$. Obviously, if $\Adversary$ can generate a false zero-knowledge proof $\pi^*_{i\,t^*}$ for $\bar{x}_{i\,t^*}$, so we can make use of $\Adversary$ to break the simulation-extractability of $\NIZK$.

\vspace{1mm}\noindent\textbf{Proof of Theorem~\ref{thm:pptp_anonymity_baseline}.} 
We first reduce the security to that of the pseudo-random function $\PRF$ by replacing the slot secret $ r_{j^*\,t^*}$ (which is supposed to compute the challenge evidence $\Evidence_{t^*}$) with a uniform random value. 
If there is an adversary $\Adversary$ which can distinguish this modification, then we could make use of $\Adversary$ to build an efficient algorithm $\mathcal{B}$ to break the security of $\PRF$. 
Since the slot secret $ r_{j^*\,t^*}$ is uniform random now, this can enable us to reduce the security to the hiding property of commitment scheme. Namely, if there exists an adversary $\Adversary$ which can distinguish the bit $b$ in the privacy game, we can build an efficient algorithm $\mathcal{B}$ running $\Adversary$ to break the hiding property of the commitment scheme. 
$\mathcal{B}$ can return the bit $b’$ obtained from $\Adversary$ to the commitment-challenger to win the game. 
Now, we use a simulator of $\NIZK$ to generate the zero-knowledge proof for challenge measurements $( x^0_{j^*\,t^*}, x^1_{j^*\,t^*})$ without using the corresponding witness. 
Similarly, any adversary which distinguishes this change, can be used to break the zero-knowledge property of $\NIZK$. 
Since the $\NIZK$ does not leak any information of the committed measurement, so we can always use $x^1_{j^*\,t^*}$ (which is random) to generate the challenge evidence. 
Note that from the compromised slot secrets at the time $t^*$, the adversary can learn at most $n-1$ measurements. But our scheme does not reveal the sum the measurements, so the adversary cannot directly get the measurement used for computing $c_{j^*\,t^*}$ and $\pi_{j^*\,t^*}$. 
Namely, we change the challenge evidence to be one which is independent of the bit $b$, so the advantage of the adversary is zero after all the above changes. 
In a nutshell, due to the security of the building blocks, the adversary can only have negligible advantage in breaking our baseline scheme.

\vspace{1mm}\noindent\textbf{Proof of Theorem~\ref{thm:pol_transparency_merkle}.} 
The commitments of leaf nodes are generated identically to those of baseline scheme. By our assumption there are at most $f$ dishonest auditors, so there must have at least one auditor's verification result (on the entire Merkle tree) among those $f+1$ signed results published on bulletin board within required time $T$ is faithful. If the Merkle tree if correctly built, then those leaf nodes' commitments are bind to the measurements from users except for a negligible property due to the binding property of the commitment scheme. Besides, the additively homomorphic property does not affect the binding property of the aggregated non-leaf nodes in the Merkle tree, which implies each commitment of the corresponding correctly commit the sum of the committed values of its children. With a similar argument in the proof of Theorem~\ref{thm:pptp_transparancy_baseline}, the zero-knowledge proofs for the non-leaf nodes would not deviate from the range of the corresponding committed value except for a negligible property because of the simulation-extractable property. Hence, we can conclude that the Merkle tree scheme achieves transparency.

\vspace{1mm}\noindent\textbf{Proof of Theorem~\ref{thm:pol_anonymity_merkle}.} 
Since the adversary can reveal at most $n-1$ slot secrets of users at the challenge time $t^*$, she can learn $n-1$ measurements of the compromised users. In this case, only the challenge measurement $x^b_{j^*\,t^*}$ and the sum of its parent node in the Merkle tree is unknown to the adversary. Note that the sub-tree involving the challenge user $j^*$ and its sibling forms is identical to the baseline scheme with just two users. So if there exists an adversary $\Adversary$ who can breaks the privacy of the baseline scheme with users $n'=2$, then we can make use of it to build an efficient algorithm $\mathcal{B}$ to break the privacy of the Merkle tree scheme. It is not hard to see that $\mathcal{B}$ can forward the challenge measurements from $\Adversary$ to its privacy-challenger to receive back the challenge evidence for simulating the challenge response to $\Adversary$. All other queries from $\Adversary$ can be simulated based on the secrets chosen by $\mathcal{B}$. In a nutshell, the privacy of the Merkle tree scheme is implied by that of the baseline scheme. 
\end{document}